\documentclass[epsfig]{kluwer}
\usepackage{epsfig}

\begin{document}
\begin{article}

\begin{opening}

\title{ISO Images of Starbursts and Active Galaxies}

\author{I.F. \surname{Mirabel$^{1,2}$ \& O. Laurent$^1$}}
\institute{\vspace{-0.5cm}}
\institute{$^1$CEA/DSM/DAPNIA, Service d'Astrophysique, 91191 Gif/Yvette. France\\
$^2$IAFE. cc 67, suc 28. Ciudad Universitaria. 1428 Buenos Aires. Argentina\\}

\runningtitle{ISO images of Starbursts and AGNs}
\runningauthor{I.F. Mirabel}

\begin{abstract} 
We present some highlights from the mid-infrared (5-16 $\mu$m) images
of mergers of massive galaxies obtained with the Infrared Space Observatory
(ISO).  We have observed: 1) ultraluminous infrared nuclei, 2)
luminous dust-enshrouded extranuclear starbursts, and 3) active galaxy
nuclei (AGNs).  In this contribution we discuss the observations of
Arp 299, a prototype for very luminous infrared galaxies, the Antennae
which is a prototype of mergers, and Centaurus A which is the closest
AGN to Earth. From these observations we conclude the following: 1)
the most intense starbursts in colliding systems of galaxies and the
most massive stars are dust-enshrouded in regions that appear
inconspicuous at optical wavelengths, 2) the most intense nuclear
infrared sources are a combination of AGN and starburst activity, 3)
the hosts of radio loud AGNs that trigger giant double-lobe structures
may be symbiotic galaxies composed of barred spirals inside
ellipticals.


\end{abstract}
\end{opening}

\vspace{-0.2cm}
\section{Nuclear Starbursts}
\vspace{-0.2cm}

The starbursts in nearby ultraluminous galaxies (\citeauthor{Sanders})
take place primarily in the nuclear regions. Using ISOCAM we have made
observations of the mid-infrared emission at 5-16 $\mu$m in a sample
of 10 very luminous galaxies. At a distance of 42 Mpc, Arp 299 (Mrk
171; NGC 3690/IC 694) is the closest system of this class with a
bolometric luminosity of 8$\times$10$^{11}$L$_{\odot}$. The nuclei are 
still $\sim$5 kpc apart, and it represents a merger in an earlier 
stage of evolution compared with  
NGC 6240 and Arp 220, where the nuclei are $\leq$1 kpc apart. 
The upper panel of Figure 1 (from Hibbard \& Yun, private 
communication) exhibits optical and HI tidal
tails 160 kpc in length emerging from the colliding disks. It is
striking that the HI tail is spatially displaced from the optical
tail.  The lower panel shows the 7 $\mu$m emission from \citeauthor{Gallais}. 
The 7 and 15 $\mu$m images reveal that about 90\% of the
emission from the whole system comes from the two unresolved
dust-enshrouded sources A and B1 which have sizes of less than 300 pc
in radius and are inconspicuous in the optical. These two regions are
strong HCN sources with some indication of rotation in source B1
\citeauthor{Casoli}. Although the studies of Arp 299 at other
wavelengths support the hypothesis that A and B1 are nuclear
starbursts, the CVF mid-infrared spectrum of B1 exhibits -besides 
the signatures of starbursts- a 3-10 $\mu$m continuum frequently 
observed in AGNs \citeauthor{Laurent}. 

\begin{figure}
\hspace{-3mm}
\centerline{\epsfig{file=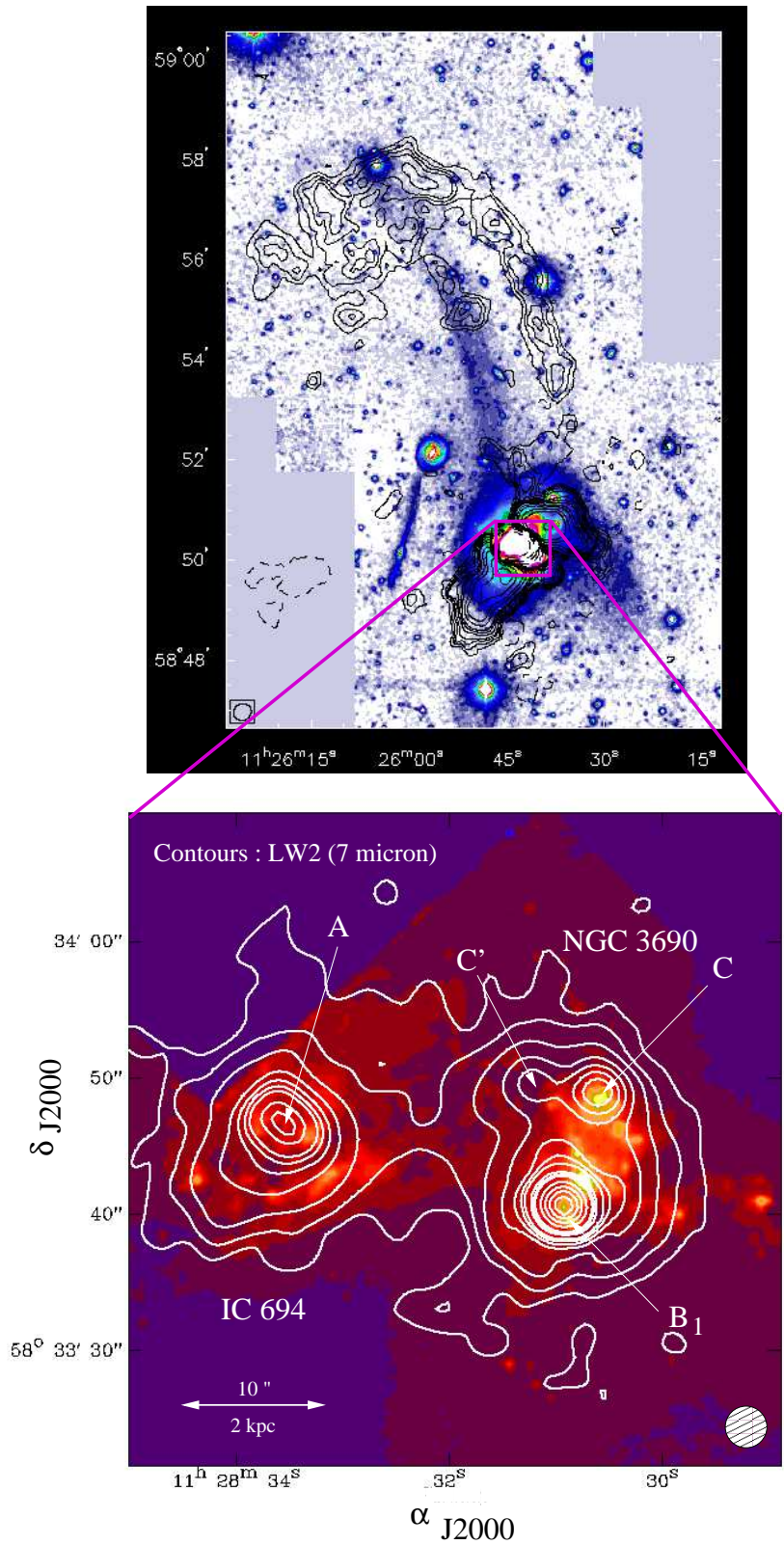,width=12cm}}
\vspace{-5mm}
\caption
{The upper figure shows R band (grey) and HI (contours) images of Arp 299 
(Hibbard \& Yun, private communication). The 7$\mu$m map by 
Gallais et al. (1999) is shown in the lower panel superposed to the optical 
HST image. About 90\% of the mid-infrared emission comes from the nuclear regions A and B1.}
\end{figure}

Source A dominates in the far infrared whereas source B in the mid-infrared 
\citeauthor{Joy}, and it can be 
concluded that more than 90\% of the bolometric luminosity in Arp 299 
comes from the 
two nuclei with sizes $\leq$300pc in radius. Similar results were
obtained for ultraluminous infrared galaxies (e.g. The Superantennae,
\citeauthor{Mirabel}), where more than 98\% of the mid-infrared emission at
15$\mu$m comes from a nuclear region hosting a Seyfert 2 nucleus.

\vspace{-0.2cm}
\section{Extranuclear Starbursts}
\vspace{-0.2cm}

One of the new findings with ISO is a very luminous
dust-enshrouded\, extranuclear\, starburst\, in the Antennae (NGC 4038/39).  
In this early merger of two Sc galaxies 
we found an extranuclear starburst with size $\leq$ 50 pc in radius  
producing $\sim$15\% of the overall 15 $\mu$m mid-infrared output. 
Furthermore, the analyses of the mid-infrared spectra
indicate that the most massive stars in this system are formed
inside this optically invisible knot.

\begin{figure} \hspace{-1mm}
\centerline{\epsfig{file=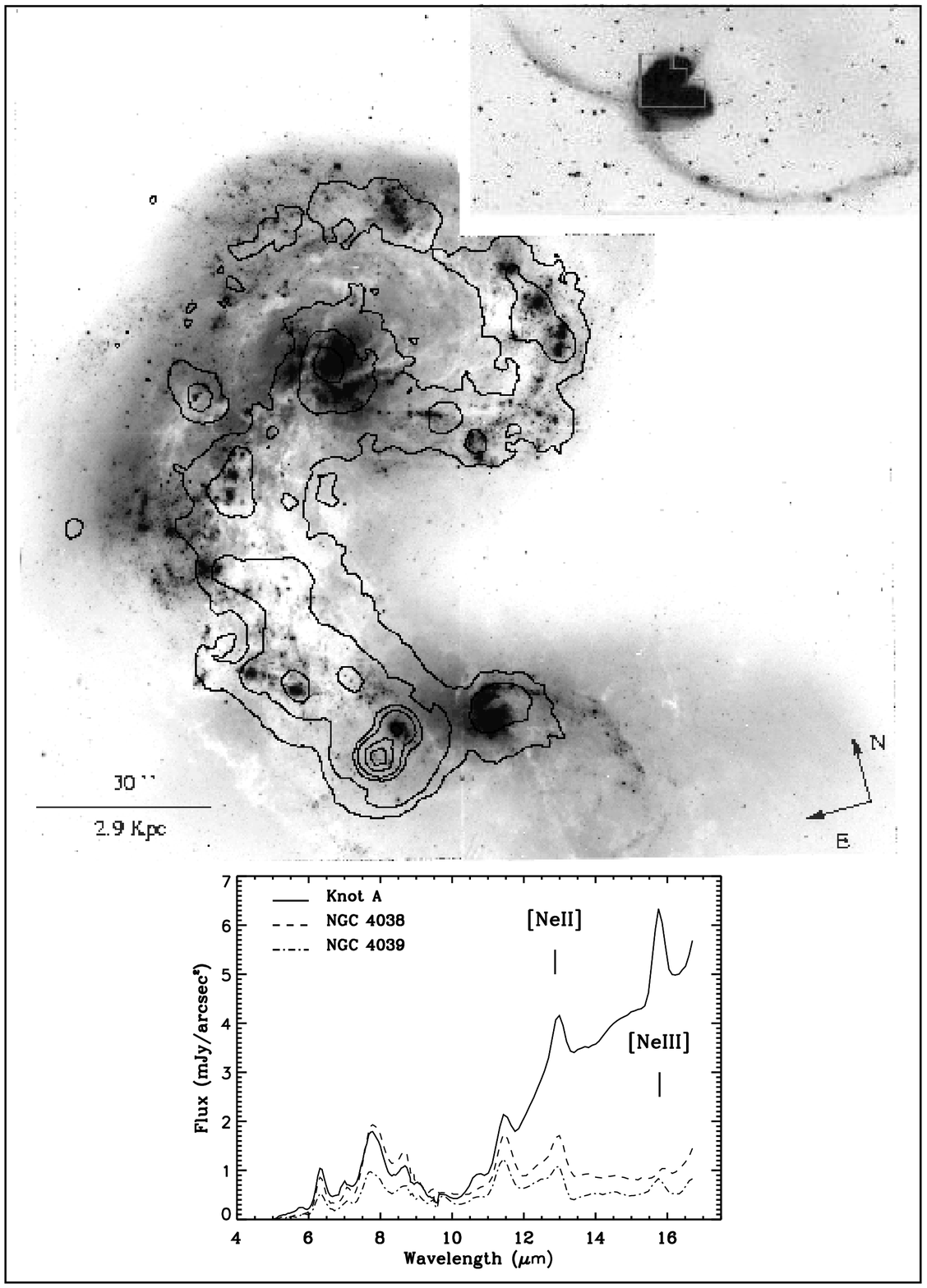,width=12cm}} \caption
{The upper figure from Mirabel et al. (1998) shows a superposition of
the mid-infrared (12 -18$\mu$m, contours) image of the Antennae
galaxies obtained with the Infrared Space Observatory, on the
composite optical image with V (5252 \AA) and I (8269 \AA) filters
recovered from the Hubble Space Telescope archive .  About half of the
mid-infrared emission from the gas and dust that is being heated by
recently formed massive stars comes from an off-nuclear region that is
clearly displaced from the most prominent dark lanes seen in the
optical. The brightest mid-infrared emission comes from a region that
is relatively inconspicuous at optical wavelengths. The ISOCAM image
was made with a 1.5$''$ pixel field of view. Contours are 0.4, 1, 3,
5, 10, and 15 mJy.  The lower figure shows the spectrum of the
brightest mid-infrared knot (continuous line) and of the 50 pc radius 
regions containing the nuclei of 
NGC 4038 and NGC 4039. The rise of the continuum above 10 $\mu$m and strong NeIII line
emission observed in the brightest mid-infrared knot indicate that the
most massive stars in this system of interacting galaxies are being
formed in that optically obscured region, still enshrouded in large
quantities of gas and dust.}
\end{figure}

In Figure 2 is shown in contours the mid-infrared (12-18 $\mu$m) image
of the Antennae galaxies obtained with ISO (\citeauthor{Mirabela}),
superimposed on the optical image from HST. Below are shown
representative spectra of the two nuclei and the brightest
mid-infrared knot. 

The multiwavelength view of this prototype merging system suggests
caution in deriving scenarios of early evolution of galaxies at high
redshift using only observations in the narrow rest-frame ultraviolet
wavelength range (Mirabel et al. 1998). Although the actual numbers of
this type of systems is not large in the local universe, we must keep
in mind that there are indications of strong number density evolution 
as a function
of redshift in luminous infrared galaxies 
(\citeauthor{Sanders}; \citeauthor{Kim}),
that the most intense starbursts are enshrouded in dust, and that no
ultraviolet light leaks out from these regions.
\\ 

Another example of an extra-nuclear starburst is also observed in the
prototypical collisional ring galaxy ``The Cartwheel'' (\citeauthor{Charmandaris}). In Figure 3 is shown in contours the 
mid-infrared image
in broad band filters LW2(5-8.5$\mu$m) and LW3(12-18$\mu$m). The
mid-IR maps cover the Cartwheel galaxy and the two nearby companions
G1 and G2. The intensity of the mid-IR emission from the outer star
forming ring of the Cartwheel shows considerable azimuthal variation
and peaks at the most active H$\alpha$ regions of the
ring. Interestingly, in the LW3(12-18$\mu$m) filter, only the hot-spot
is detected in the outer ring. In addition, the LW3/LW2 flux ratio,
often used as a diagnostic of the intensity of the radiation field has a value of 5.2 which is comparable to the brightest extranuclear starburst in the Antennae galaxies.

\begin{figure}
\hspace{-1mm}
\centerline{\epsfig{file=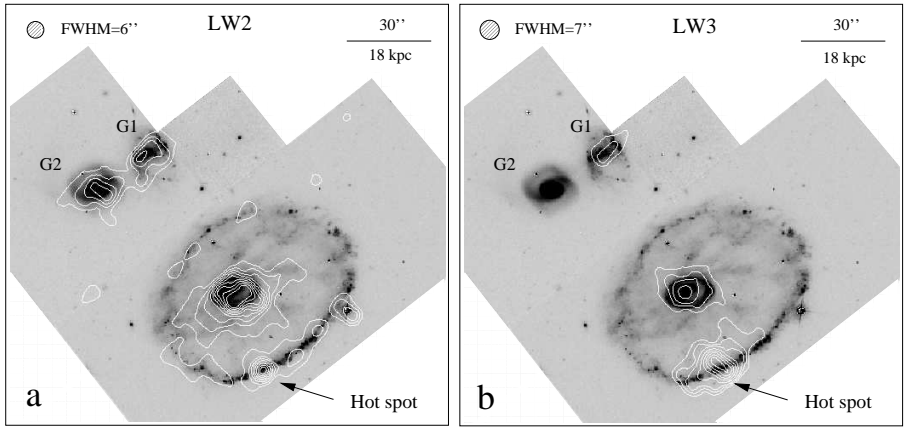,width=12cm}}
\caption{
a) Contour map of the ISOCAM LW2 emission overlaid on an HST
wide I band image (Charmandaris et al. 1999). The contour levels are from 
0.1 to 0.5 mJy/pixel with a
step of 0.05 mJy/pixel.  b) Contour map of the ISOCAM LW3 emission
overlaid on the same HST image. The contour levels are
0.2,0.3,0.4,0.6,0.8,1,1.2 and 1.4 mJy/pixel. North is up and East is 
left in both images.
}
\end{figure}

\vspace{-0.4cm}
\section{AGNs and Symbiotic Galaxies}
\vspace{-0.3cm}

\begin{figure}
\hspace{-1mm}
\centerline{\epsfig{file=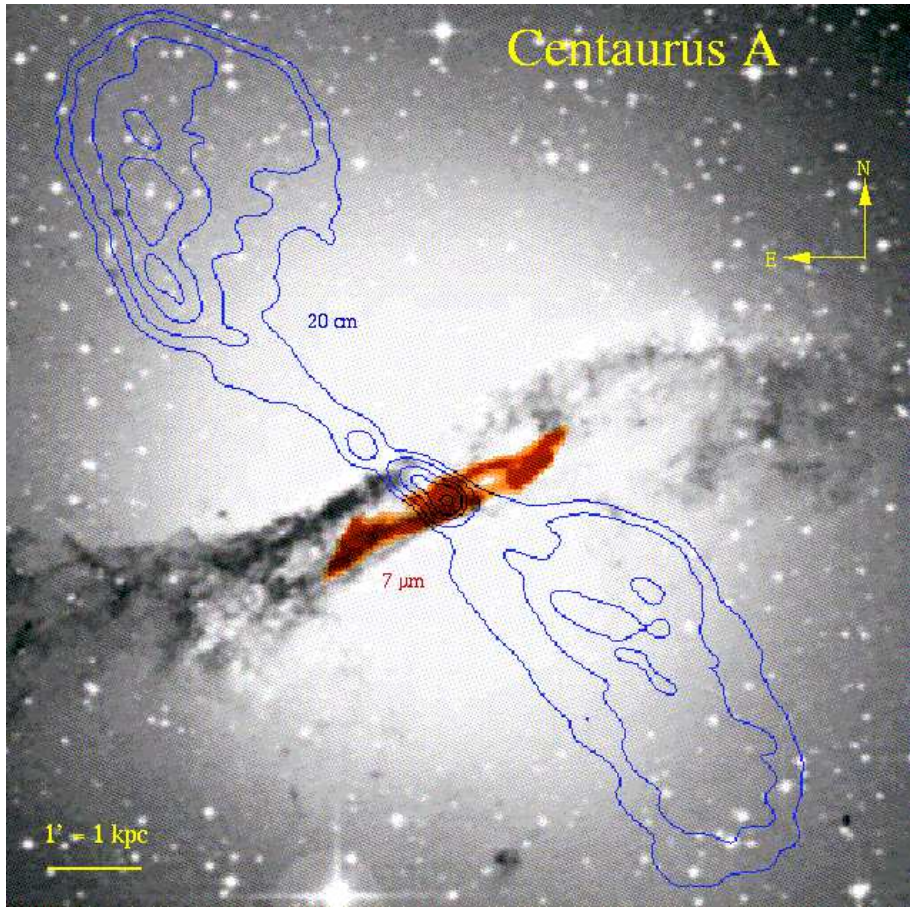,width=12cm}}
\caption
{The ISO 7\,$\mu$m emission (dark structure; Mirabel et al. 1999) and
VLA 20 cm continuum in contours (Condon et al. 1996), overlaid on an
optical image from the Palomar Digital Sky Survey. The 7 $\mu$m emission from
dust with a bisymmetric morphology at the centre is about 10 times
smaller than the overall size of the shell structure in the elliptical
and lies on a plane that is almost parallel to the
minor axis of its giant host. Whereas the gas associated with the spiral
rotates with a maximum radial velocity of 250 km s$^{-1}$, the
ellipsoidal stellar component rotates slowly approximately
perpendicular to the dust lane (Wilkinson et al. 1986). The
synchrotron radio jets shown in this figure correspond to the inner
structure of a double lobe radio source that extends up to 5$^{\circ}$
($\sim$ 300 kpc) on the sky. The jets are believed to be powered by a
massive black hole located at the common dynamic center of the
elliptical and spiral structures.}
\end{figure}

Giant radio galaxies are thought to be massive ellipticals powered by
accretion of interstellar matter onto a supermassive black
hole. Interactions with gas rich galaxies may provide the interstellar
matter to feed the active galactic nucleus (AGN). To power radio lobes
that extend up to distances of hundreds of kiloparsecs, gas has to be
funneled from kiloparsec size scales down to the AGN at rates of
$\sim$1 M$_{\odot}$ yr$^{-1}$ during $\sim$10$^8$ years. Therefore,
large and massive quasi-stable structures of gas and dust should exist
in the deep interior of the giant elliptical hosts of double lobe
radio galaxies. Recent mid-infrared observations with ISO revealed for
the first time a bisymmetric spiral structure with the dimensions of a
small galaxy at the centre of Centaurus A (Mirabel et al. 1999). The
spiral was presumably formed out of the tidal debris of an accreted gas-rich
object(s) and has a dust morphology that is remarkably similar to that
found in barred spiral galaxies (see Figure 4). The observations of
this closest AGN to Earth suggest that the dusty hosts of giant radio
galaxies like CenA, are ``symbiotic" galaxies composed of a barred
spiral inside an elliptical, where the bar serves to funnel gas toward
the AGN.

The barred spiral at the centre of CenA has dimensions comparable to
that of the small Local Group galaxy Messier 33. It lies on a plane
that is almost parallel to the minor axis of the giant
elliptical. Whereas the spiral rotates with maximum radial velocities
of $\sim$\,250 km s$^{-1}$, the ellipsoidal stellar component seems
to rotate slowly (maximum line-of-sight velocity is $\sim$\,40 km
s$^{-1}$) approximately perpendicular to the dust lane. The genesis,
morphology, and dynamics of the spiral formed at the centre of CenA
are determined by the gravitational potential of the elliptical, much
as a usual spiral with its dark matter halo. On the other hand, the
AGN that powers the radio jets is fed by gas funneled to the center
via the bar structure of the spiral. The spatial co-existence and
intimate association between these two distinct and dissimilar systems
suggest that Cen A is the result from a cosmic symbiosis.

\acknowledgements
We are grateful to J. Hibbard, P. Gallais, and V. Charmandaris for 
permission to show figures 1 and 3.  

{}

\end{article}
\end{document}